# Diffusion MRI with double diffusion encoding and variable mixing times disentangles water exchange from intrinsic kurtosis

*Arthur Chakwizira[1], Filip Szczepankiewicz[1], and Markus Nilsson[2]

1. Medical Radiation Physics, Clinical Sciences Lund, Lund University, Lund, Sweden
2. Department of Clinical Sciences Lund, Radiology, Lund University, Lund, Sweden

## Corresponding author:

Arthur Chakwizira

Department of Medical Radiation Physics, Lund University, Skåne University Hospital,

SE-22185 Lund, Sweden

Email address: arthur.chakwizira@med.lu.se

## Word count: 7000

## Sponsors/Grant numbers:

- VR (Swedish Research Council)
    - 2020-04549
    - 2021-04844
- eSSENCE
    - 10:5
- Cancerfonden (The Swedish Cancer Society)
    - 2022/2414
    - 22 0592 JIA

## Abstract

Double diffusion encoding (DDE) makes diffusion MRI sensitive to a wide range of microstructural features, and the acquired data can be analysed using different approaches. Correlation tensor imaging (CTI) uses DDE to resolve three components of the diffusional kurtosis: isotropic, anisotropic, and microscopic. The microscopic kurtosis is estimated from the contrast between single diffusion encoding (SDE) and parallel DDE signals at the same b-value. Another approach is multi-Gaussian exchange (MGE), which employs DDE to measure exchange. Sensitivity to exchange is obtained by contrasting SDE and DDE signals at the same b-value. CTI and MGE exploit the same signal contrast to quantify microscopic kurtosis and exchange, and this study investigates the interplay between these two quantities. We perform Monte-Carlo simulations in different geometries with varying levels of exchange and study the behaviour of the parameters from CTI and MGE. We conclude that microscopic kurtosis from CTI is sensitive to the exchange rate and that intercompartmental exchange and the intrinsic kurtosis of individual compartments are distinct sources of microscopic kurtosis. In an attempt to disentangle these two sources, we propose a heuristic signal representation referred to as µMGE (MGE incorporating intrinsic kurtosis) that accounts for both effects, by exploiting the distinct signatures of exchange and intrinsic kurtosis with varying mixing time: exchange causes a slow dependence of the signal on mixing time while intrinsic kurtosis arguably has a much faster dependence. We find that applying µMGE to data acquired with multiple mixing times for both parallel and orthogonal DDE may allow estimation of the exchange rate as well as the isotropic, anisotropic, and intrinsic kurtosis.

## 1 Introduction

Diffusion magnetic resonance imaging (dMRI) is a powerful radiological tool due to both its non-invasive nature and its exquisite sensitivity to the microstructure of biological tissue [1–3]. It provides significant value to radiology, for example, by enabling measurements of the apparent diffusion coefficient (ADC). The utility of the ADC is its high sensitivity to microstructural tissue changes [4–9]. However, it is not very specific because it responds to multiple aspects of tissue microstructure [10–12]. A key reason for the poor specificity of the ADC, and dMRI in general, is the experimental design on which most dMRI applications rely: the so-called single diffusion encoding (SDE) scheme proposed in 1965 by Stejskal and Tanner [13] and later designated this name by Shemesh et al. [14]. Although widely used for a variety of purposes [4,15–28], SDE is an insufficient probe of heterogenous tissue because it conflates several microstructural features such as microscopic diffusion anisotropy, orientation dispersion and isotropic heterogeneity [29–31]. To address the shortcomings of SDE, the double diffusion encoding (DDE) scheme, which uses two pairs of diffusion-sensitising pulses separated by a mixing time was introduced in a pioneering study in 1990 [32]. That study showed that DDE acquisitions with parallel and orthogonal gradient pairs enabled measurement of the local eccentricity of a sample, even when the sample appeared isotropic at the voxel scale due to orientation dispersion. It was later shown that additional specificity can be gained by DDE in other scenarios too, for example, to separate microscopic anisotropy from orientation dispersion [29,33–35], measure flow [36,37], estimate pore sizes [38,39], probe microscopic kurtosis [40,41] and to measure exchange [42–47].

Correlation tensor imaging (CTI) was recently proposed as a means of using DDE to measure the so-called intra-compartmental or microscopic kurtosis [40,41]. The approach is based on the cumulant expansion of the DDE signal [48] and it assumes the long-mixing-time regime where the displacement correlation tensor is proportional to the diffusion covariance tensor [35,49] – a relation that allows estimation of isotropic and anisotropic kurtosis [30,31,40]. In practice, the microscopic kurtosis is obtained by subtracting the anisotropic and isotropic kurtoses from the total kurtosis. Notably, the information used to estimate microscopic kurtosis in CTI is the contrast between an SDE and a parallel DDE acquisition with the same total b-value [41].

DDE can also be used to measure exchange, by using variable mixing times, as proposed in conjunction with both spectroscopy [43,44] and imaging [45,46,50]. In so-called filter-exchange imaging (FEXI), the first pair of diffusion-sensitising gradients (the filter block) is used to suppress the contribution of fast-diffusing water to the measured signal, leading to a decrease in the ADC [45,46]. The return to equilibrium is tracked by measuring the ADC as a function of the mixing time, and the rate of this equilibration is called the apparent exchange rate (AXR). More recent work has unified all experimental designs for probing exchange with dMRI by defining for a given gradient waveform an "exchange sensitivity"[47,51]. The approach is derived assuming multi-Gaussian diffusion and slow-to-intermediate exchange rates. Here, we refer to this framework as multi-Gaussian exchange (MGE), wherein the highest sensitivity to exchange is obtained by contrasting SDE and (parallel) DDE acquisitions with the same b-value.

CTI and MGE both exploit the contrast between SDE and parallel DDE acquisitions at a fixed b-value: the former attributes it to microscopic kurtosis and the latter to exchange. Previous work has highlighted that exchange may be one of the sources of the microscopic kurtosis estimated by CTI [40,41,52]. This work aims to investigate how exchange affects estimates of the microscopic kurtosis from CTI and the exchange rate from MGE by running Monte-Carlo simulations in different substrates with varying levels of exchange. This work also attempts to separate exchange from the other sources of microscopic kurtosis (the other sources are herein referred to as "intrinsic kurtosis"). This is achieved by first extending the MGE theory to account for exchange in the presence of anisotropy, enabling the simultaneous estimation of isotropic and anisotropic kurtosis sources as well as exchange. The extended MGE theory also incorporates persistent exchange-independent kurtosis sources that may result from residual voxel anisotropy or powder averaging. Intrinsic kurtosis is then introduced as an additional exchange-independent source of diffusional kurtosis. What enables separation of intrinsic kurtosis from exchange in this framework is the distinct signatures that the two processes have on the signal with varying mixing time: exchange causes a dependence on mixing time while intrinsic kurtosis (in the long mixing time regime) does not. We interpret our results as an indication that exchange and intrinsic kurtosis can be disentangled by using the unified approach described above (μMGE).

## 2 Theory

*The cumulant expansion*

The diffusion-weighted MRI signal can be expressed as the Laplace transform of the spin phase distribution [53,54]

$$E = S/S_0 = \int P(\varphi)d\varphi = \langle \exp(-i\varphi) \rangle, \qquad (1)$$

where the averaging $\langle \cdot \rangle$ is done over all contributing spins in the voxel, $E$ is the normalized signal, $S_0$ is the non-diffusion-weighted signal and $\varphi$ is the phase defined through the spin trajectory $\boldsymbol{r}(t)$ and the diffusion-encoding gradient waveform $\boldsymbol{g}(t)$ as

$$\varphi = \gamma \int_0^T \boldsymbol{g}(t) \cdot \boldsymbol{r}(t) \, dt, \tag{2}$$

where $\gamma$ is the gyromagnetic ratio and $T$ is the total duration of $\boldsymbol{g}(t)$. Equation 1 can be approximated by taking its cumulant expansion

$$\ln(E) \approx -\frac{1}{2}\langle \varphi^2 \rangle + \frac{1}{24}(\langle \varphi^4 \rangle - 3\langle \varphi^2 \rangle^2) = -\frac{1}{2}c_2 + \frac{1}{24}c_4, \tag{3}$$

where $c_2$ and $c_4$ are the second and fourth cumulants of the distribution of $\varphi$. In heterogenous media comprising multiple local environments with distinct diffusion properties, averaging must be performed both over spins in each environment and over the environments themselves. Let $o_2$ and $o_4$ denote the environment-specific second and fourth cumulants. It can be shown that [48,51,55]

$$c_2 = \langle o_2 \rangle_e \tag{4}$$

and

$$c_4 = \langle o_4 \rangle_e + 3(\langle o_2^2 \rangle_e - \langle o_2 \rangle_e^2) \tag{5}$$

where the subscript "e" denotes averaging over environments. Note that, fundamentally, the signal is given by an average over spins, and this can be divided into an average over environments plus an average over spins within them when the exchange between the environments is negligible during the time $T$. The first term of Eq. 5 captures the intra-compartmental or intrinsic variance (or kurtosis if $c_4$ is normalized by $c_2^2$) while the second term is the inter-compartmental variance which may incorporate both isotropic and anisotropic components. Note that $\langle o_4 \rangle_e$ is zero for Gaussian diffusion but for restricted diffusion it is negative or positive depending on the diffusion time[20]. The negative case can be understood by considering the signal-versus-b curve of such environments. For low b-values, it exhibits a mono-exponential attenuation in line with the Gaussian phase approximation [56,57]. At higher b-values, it declines more rapidly as the famous diffraction pattern of restricted samples begins to form [58–60]. This super-exponential attenuation yields negative kurtosis. In more complex geometries, however, $\langle o_4 \rangle_e$ could be positive even though diffusion is restricted due to effects of cross-sectional variance. An analysis of how cross-sectional variance attenuates diffraction patterns was previously reported [61].

*CTI*

Correlation tensor imaging leverages a combination of SDE, parallel DDE and orthogonal DDE measurements in the long-mixing-time regime to resolve three components of $c_4$: isotropic, anisotropic, and microscopic kurtosis [40,41]. Assuming powder-averaging [35,62], also known as spherical averaging [63], Eq. 3 evaluates in CTI to

$$\ln(\bar{E}_{DDE}(b_1, b_2, \theta))$$
$$\approx -(b_1 + b_2)\bar{D} + \frac{1}{6}(b_1^2 + b_2^2)\bar{D}^2 K_T + \frac{1}{2}b_1 b_2 \cos^2(\theta)\bar{D}^2 K_A$$
$$+ \frac{1}{6}b_1 b_2 \bar{D}^2 (2K_I - K_A), \qquad (6)$$

where $b_1$ and $b_2$ are the b-values of the first and second gradient pairs, $\theta$ is the angle between the two pairs, $\bar{D}$ is the mean diffusivity, $K_T$ is the total kurtosis which is a sum of isotropic ($K_I$), anisotropic ($K_A$) and microscopic ($K_\mu$) contributions, such that

$$K_\mu = K_T - K_I - K_A. \qquad (7)$$

Note that $K_\mu$ is negative for restricted diffusion in simple geometries [20,40]. For example, diffusion restricted in a sphere exhibits no $K_I$ and $K_A$, and thus $K_T = K_\mu$.

An alternative representation can be obtained by defining distinct quantities that are sensitive to the different kurtosis components. To begin with, the shape of the encoding b-tensor can be defined as [31,64]

$$b_\Delta^2 = \frac{b_1^2 + b_2^2 + b_1 b_2 (3\cos^2\theta - 1)}{(b_1 + b_2)^2} = \begin{cases} 1; & \text{SDE or parallel DDE} \\ \frac{1}{4}; & \text{orthogonal DDE if } b_1 = b_2 \end{cases} \qquad (8)$$

Furthermore, we define a new metric that reports on how sensitive the encoding is to the microscopic kurtosis, following the same notion as above, according to

$$b_\mu^2 = \frac{b_1^2 + b_2^2}{(b_1 + b_2)^2} = \begin{cases} 1; & \text{SDE} \\ \frac{1}{2}; & \text{DDE if } b_1 = b_2 \end{cases} \tag{9}$$

Using these metrics, we can now rewrite the central CTI equation according to

$$\ln\left(\bar{E}_{DDE}(b, b_\Delta^2, b_\mu^2)\right) = -b\bar{D} + \frac{1}{6}b^2\bar{D}^2(K_I + b_\Delta^2 K_A + b_\mu^2 K_\mu) \tag{10}$$

where $b = b_1 + b_2$. Thus, in the CTI framework, sensitivity to $K_A$ is obtained by contrasting parallel and orthogonal DDE measurements, namely

$$\ln\left(\bar{E}_{DDE}\left(b, 1, \tfrac{1}{2}\right)\right) - \log\left(\bar{E}_{DDE}\left(b, \tfrac{1}{4}, \tfrac{1}{2}\right)\right) = \frac{1}{8}b^2\bar{D}^2 K_A \tag{11}$$

and sensitivity to $K_\mu$ is obtained by contrasting SDE and parallel DDE measurements

$$\ln\left(\bar{E}_{DDE}(b, 1, 1)\right) - \ln\left(\bar{E}_{DDE}\left(b, 1, \tfrac{1}{2}\right)\right) = \frac{1}{12}b^2\bar{D}^2 K_\mu \tag{12}$$

The microscopic kurtosis assumes measurements in the long mixing time limit, where $K_\mu$ is insensitive to the mixing time, and where the displacement correlation tensor becomes proportional to the diffusion covariance tensor. The long mixing time limit has in previous works been defined as fulfilled when there is a vanishing difference between parallel and antiparallel DDE signals [41].

*MGE in one dimension*

The one-dimensional multi-Gaussian exchange framework (1D-MGE) assumes negligible intra-compartmental kurtosis but incorporates exchange between Gaussian environments [47,51]. In this framework, Eq. 3 for powder-averaged DDE signals ($\bar{E}_{DDE}$) evaluates to the following

$$\ln(\bar{E}_{DDE}(b, h(k))) \approx -b\bar{D} + \frac{1}{6}b^2\bar{D}^2 K_T h(k), \tag{13}$$

where $k$ is the intercompartmental exchange rate and $h(k)$ is the exchange-weighting function given by

$$h(k) = 2\int_0^T \tilde{q}_4(t) \exp(-kt)\,dt, \tag{14}$$

where $\tilde{q}_4(t) = q_4(t)/b^2$ and $q_4(t)$ is the fourth-order autocorrelation function of the dephasing q-vector given by $q_4(t) = \int_0^T q^2(t')q^2(t'+t)dt'$. This approach assumes the diffusional heterogeneity of a system to be described by $K_T$ and its homogenization over time by $h(k)$. Waveforms are more sensitive to the homogenization (exchange) when they feature strong phase-dispersion power ($q^2$) separated in time, as exchange means that, with time, the diffusivity becomes less correlated, which decreases the observed heterogeneity, $K_T h(k)$. Note that this theory was developed assuming two exchanging components. Exchange in multiple independent systems could be analysed by applying Eq. 5 to this setup, meaning Eq. 14 would become multiexponential. In this work, we will investigate the condition where there are two types of systems, either in exchange with the same exchange rate, or not in exchange at all. An analysis of systems exhibiting multiple distinct exchange rates is beyond the scope of this report, but a theoretical starting point for such an analysis can be found in [65].

For SDE with narrow pulses, the function $h$ evaluates to [47]

$$h_{SDE}(k, \Delta) = \frac{2}{k\Delta} - \frac{2}{(k\Delta)^2} + \frac{2}{(k\Delta)^2} e^{-k\Delta}, \quad (15)$$

where $\Delta$ is the pulse spacing. Correspondingly for short-pulse DDE assuming for simplicity $b_1 = b_2$,

$$h_{DDE}(k, \Delta, t_m) = \frac{1}{2} h_{SDE}(k, \Delta) + \frac{1}{2(k\Delta)^2} \left(e^{-kt_m} + e^{-k(2\Delta + t_m)} - 2e^{-k(\Delta + t_m)}\right), \quad (16)$$

where $t_m$ is the mixing time and we have assumed $\Delta_1 = \Delta_2 = \Delta$. Expressions for the exchange weighting function for acquisitions with finite pulse widths can be found in [47].

Notably, sensitivity to the exchange rate in 1D-MGE can be obtained by contrasting SDE and parallel DDE signals at a fixed b-value

$$\ln(\bar{E}_{DDE}(b, h_{SDE})) - \ln(\bar{E}_{DDE}(b, h_{DDE})) = \frac{1}{6} b^2 \bar{D}^2 K_T \cdot (h_{SDE} - h_{DDE}). \quad (17)$$

The 1D-MGE framework assumes effects of restricted diffusion can be neglected. Acquisition protocols can be designed to partially support such an assumption, by featuring waveforms with identical sensitivity to restricted diffusion [51]. The DDE-based filter exchange imaging (FEXI) approach is an example of such a protocol [45].

*Comparing 1D-MGE and CTI*

The contrast between SDE and parallel DDE at a fixed b-value is captured by different parameters in CTI and 1D-MGE: $K_\mu$ in the former and $k$ in the latter. A CTI measurement on a system well-described by multiple exchanging Gaussian components will thus yield a non-zero $K_\mu$ that is driven by exchange. The outcome of such a measurement can be predicted by comparing the SDE-DDE signal contrasts from both CTI and 1D-MGE. Equating the right-hand-sides of Eq. 12 and Eq. 17 provides

$$\frac{1}{12} b^2 \overline{D}^2 K_\mu = \frac{1}{6} b^2 \overline{D}^2 K_T \cdot [h_{SDE}(k,\Delta) - h_{DDE}(k,\Delta,t_m)]. \tag{18}$$

Equivalently,

$$K_\mu = 2 K_T \cdot [h_{SDE}(k,\Delta) - h_{DDE}(k,\Delta,t_m)]. \tag{19}$$

Intuition about Eq. 19 can be gained by assuming $k\Delta \ll 1$ and $kt_m \ll 1$ and approximating the exponential terms in $h_{SDE}$ and $h_{DDE}$ with third-degree polynomials. This provides

$$K_\mu = K_T \left(\frac{2}{3}\Delta + t_m\right) k. \tag{20}$$

Thus, CTI-estimated microscopic kurtosis (at finite mixing times) in a multi-Gaussian exchange setting increases with the exchange rate, mixing time and diffusion time. However, note that the assumption $kt_m \ll 1$ applied above is in principle incompatible with CTI which assumes the long mixing time regime. Equation 20 is thus a prediction of what CTI measures when the long mixing time condition is not satisfied, and thus represents an *apparent* microscopic kurtosis. The long mixing time regime can be obtained by letting $t_m$ go to infinity, in which case Eq. 19 becomes

$$K_\mu(t_m \to \infty) = K_T \cdot h_{SDE}(k,\Delta), \tag{21}$$

which, when $k\Delta \ll 1$, becomes

$$K_\mu(t_m \to \infty) = K_T \cdot \left(1 - k\frac{\Delta}{3}\right). \tag{22}$$

Therefore, even in the long mixing time regime, $K_\mu$ decreases with increasing exchange rates and diffusion time. Note that the same dependence of $K_\mu$ on exchange rate and diffusion time shown in Eq. 21 has been derived previously [52,66].

Equations 19 and 21 are illustrated in Fig. 1 for $K_T = 1$ and $\Delta = 12$ ms. At finite mixing times, $K_\mu$ increases with the exchange rate up to a $t_m$-dependent peak, after which it decreases. At infinitely long mixing times, $K_\mu$ decreases monotonically with $k$. Note that it is the product $k \cdot t_m$ that drives the approach to the long mixing time regime, thus short mixing times combined with fast exchange rates can yield the long mixing time behaviour in Eq. 19.

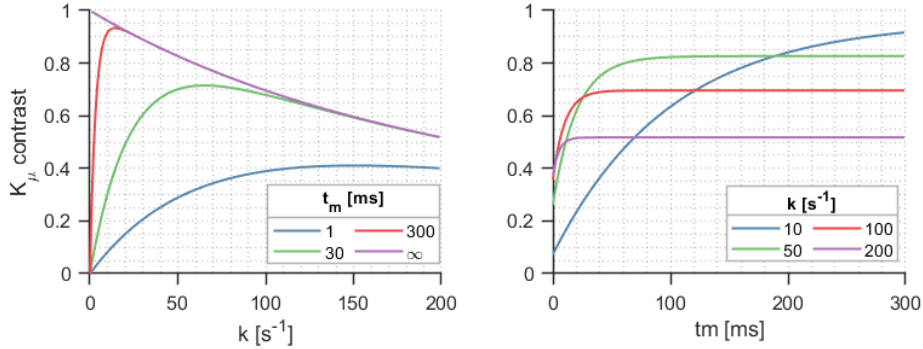

Figure 1: Dependence of the SDE-DDE signal contrast in CTI on exchange rate and mixing time. At finite mixing times, the contrast increases with exchange rate up to a mixing-time-dependent peak after which it decreases. The long mixing time regime is denoted by a monotonic decrease of the SDE-DDE contrast with exchange rate and a loss of sensitivity to further changes in the mixing time.

*Extending 1D-MGE to account for anisotropy and exchange-independent kurtosis*

The one-dimensional MGE theory presented in the previous sections describes exchange in simple systems with isotropic diffusion but cannot deal with anisotropy. We here seek to extend the theory to account for two conditions: (1) exchange in the presence of anisotropy and (2) exchange in multiple Gaussian pair-wise exchanging components where the components themselves are not in exchange. The latter condition can arise from either a system comprising such pair-wise connected components or from acquiring signals in multiple directions in a system with one or more anisotropic components. This condition gives rise to a diffusional covariance that persists independent of local exchange.

We begin by generalizing Eq. 13 to account for anisotropy, which necessitates a tensorial description according to

$$\ln E \approx -\mathbf{B} : \mathbf{D} + \frac{1}{2}\mathbb{H}(k) : \mathbb{C}^0, \qquad (23)$$

where

$$\mathbf{B} = \gamma^2 \int_0^T \mathbf{q}^{\otimes 2}(t)\, dt \qquad (24)$$

with

$$\mathbf{q}(t) = \gamma \int_0^t \mathbf{g}(t')\, dt' \qquad (25)$$

is the b-tensor [31], "$\otimes$" denotes outer tensor product, ":" denotes inner product, and $\mathbb{H}(k)$ is the exchange-sensitised "square of the b-tensor", defined as

$$\mathbb{H}(k) = 2 \int_0^T \mathbb{Q}_4(t) \exp(-kt)\, dt, \qquad (26)$$

where $\mathbb{Q}_4(t)$ is the fourth-order autocorrelation tensor of the dephasing q-vector given by

$$\mathbb{Q}_4(t) = \int_0^T \mathbf{q}^{\otimes 2}(t') \otimes \mathbf{q}^{\otimes 2}(t' + t)\, dt'. \qquad (27)$$

Note that $\mathbb{H}(k)$ becomes more weighted towards anisotropy with increasing exchange rate, as the contribution of differently oriented q-vectors at different time points, which reduces the anisotropy weighting of $\mathbb{H}(k)$, are attenuated by exchange. For example, for DDE acquisitions, as the exchange rate grows, the isotropic component of $\mathbb{H}(k)$ converges towards the anisotropic one (cf. Fig 7A). This is an interesting theoretical result, that may help explain the results studies applying filter-exchange imaging in systems with anisotropy [67,68]. Section 7.1 of the supplementary material provides more details. Note also that in the absence of exchange, $\mathbb{H}(k)$ is simply the "square of the b-tensor", that is, $\mathbb{H}(0) = \mathbf{B}^{\otimes 2}$. With exchange, it changes both in size and shape, and it is therefore not purely related to the experiment but also to the microstructure. Finally, $\mathbf{D}$ and $\mathbb{C}^0$ denote the average diffusion tensor and diffusion-tensor covariance of the local system in exchange [31].

In the presence of multiple local systems in exchange, for example in a voxel with multiple fiber orientations, we need an additional term. Recall that Eq. 4 and 5 state that the globally averaged second cumulant is simply the average of the local second cumulants, while the fourth cumulant has contributions both from the average of the local fourth cumulants, as well as the dispersion between second cumulants. Accordingly, we add terms reflecting this

$$\ln E \approx -\mathbf{B} : \langle \mathbf{D} \rangle + \tfrac{1}{2} \langle \mathbb{H}(k) : \mathbb{C}^0 \rangle + \tfrac{1}{2} \mathbf{B}^{\otimes 2} : \mathbb{C}^\infty, \tag{28}$$

where $\langle \mathbf{D} \rangle$ is the globally averaged diffusion tensor, $\mathbb{C}^\infty$ is the long-time covariance across non-exchanging ensembles that persists despite the presence of local exchange (captured by the first term of Eq. 5), given by $\mathbb{C}^\infty = \langle \mathbf{D}^{\otimes 2} \rangle - \langle \mathbf{D} \rangle^{\otimes 2}$. Thus, as time approaches infinity in the presence of exchange, $\mathbb{C}^\infty$ is the only remaining source of diffusional variance.

Assuming all exchanging systems share the same exchange rate – a strong assumption – yielding $\langle \mathbb{H}(k) : \mathbb{C}^0 \rangle = \mathbb{H}(k) : \langle \mathbb{C}^0 \rangle$, the powder average of the signal ($E$) in Eq. 28 is given by

$$\ln E \approx -bD + \frac{1}{2} b^2 h(k)[V_I^0 + h_\Delta^2(k) V_A^0] + \frac{1}{2} b^2 [V_I^\infty + b_\Delta^2 V_A^\infty] \tag{29}$$

where $b = \mathrm{Tr}(\mathbf{B})$ is the trace of the b-tensor, $h(k)$ and $h_\Delta(k)$ are the isotropic and anisotropic projections of $\mathbb{H}(k)$ given by

$$h(k) = (\mathbb{H}(k) : 9\,\mathbb{I}_\mathrm{I})/b^2 \tag{30}$$

which is identical to Eq. 14 and

$$h_\Delta^2(k) = \left(\mathbb{H}(k) : \frac{9}{2}\mathbb{I}_A\right)/(b^2 h(k)), \tag{31}$$

where $\mathbb{I}_I$ and $\mathbb{I}_A$ are fourth-order isotropic tensors capturing isotropic and anisotropic variance, respectively [31,55,64,69], $D = \text{Tr}(\langle \mathbf{D} \rangle)$ is the trace of the average diffusion tensor, $V_I^0$ and $V_A^0$ are isotropic and anisotropic variances and $V_I^\infty$ and $V_A^\infty$ are the long-time isotropic and anisotropic variances, respectively. Anisotropy of the average diffusion tensor $\langle \mathbf{D} \rangle$ will cause a non-zero $V_A^\infty$ in the powder-averaged signal.

In terms of kurtosis, Eq. 31 can be written

$$\ln E \approx -bD + \frac{1}{6}b^2 h(k) D^2 [K_I^0 + h_\Delta^2(k) K_A^0] + \frac{1}{6}b^2 D^2 [K_I^\infty + b_\Delta^2 K_A^\infty] \tag{32}$$

We refer to Eq. 32 as MGE. The kurtosis contribution $K_I^\infty$ can be non-zero due to heterogeneity in isotropic diffusivity between the different systems in exchange, and $K_A^\infty$ can be non-zero for example due to effects of residual voxel anisotropy and powder averaging, as described earlier. Note, however, that this signal representation is degenerate in the absence of exchange ($k$ = 0) because $h(0) = 1$ and $h_\Delta^2(0) = b_\Delta^2$ and thus there is no way of separating $K_I^\infty$ from $K_I^0$ or $K_A^0$ from $K_A^\infty$. This is not surprising, as separation of exchanging and non-exchanging systems requires exchange. However, this also means that the representation may become degenerate for systems in very slow exchange. Another limitation of the MGE framework is that it neglects structural disorder-induced time-dependence which is characterized by power-laws that outlive the exponential exchange-driven time-dependence [70,71].

*Extending MGE to incorporate microscopic kurtosis*

Previous sections illustrated that $K_\mu$ as measured by CTI is associated with multiple processes. For example, it is negative in environments with diffusion restricted in simple geometries, while it is positive and associated with the exchange rate in systems with exchange. Here, we propose and explore an alternative approach that may disentangle the different sources of microscopic kurtosis, by separating effects of intercompartmental exchange from the intrinsic kurtosis of local environments. We opt for the word *intrinsic* here to highlight that this quantity is not necessarily the same as microscopic kurtosis from CTI which incorporates exchange. Consider an environment featuring multiple non-exchanging non-Gaussian compartments associated with a diffusion tensor $\mathbf{D}_j$ and intrinsic kurtosis $\mathbf{W}_j$ such that the signal attenuation (powder-averaged) from a single compartment can be written

$$\ln E \approx -bD_j + \frac{1}{6}b^2 D_j^2 b_\mu^2 W_j \qquad (33)$$

Averaging over multiple such compartments results in the signal representation

$$\ln E \approx -bD + \frac{1}{6}b^2 D^2 b_\mu^2 K_{int} + \frac{1}{6}b^2 D^2 [K_I + b_\Delta^2 K_A] \qquad (34)$$

where $D = \langle D_j \rangle$, $K_{int} = \frac{\langle D_j^2 W_j \rangle}{D^2}$ and $K_I = \frac{3(\langle D_j^2 \rangle - \langle D_j \rangle^2)}{D^2}$ and $K_A = \frac{6}{5}\frac{\langle V_\lambda(\mathbf{D}_j)\rangle}{D^2}$. The above expression is in line with the premise of CTI, with $K_{int} = K_\mu$. We now consider *slow* exchange between the compartments above – slow enough for the picture of distinct compartments to remain a good representation of the system. In practical terms, we require that spins spend most of the diffusion encoding time in the same compartment, which is possible when the diffusion correlation time of the compartment is much shorter than the exchange time. This is also known as barrier-limited exchange. Under these conditions, despite the occasional excursions of spins across compartments, the quantity $K_{int}$ exists and represents the average intrinsic kurtosis of the system. The effect of the excursions is, however, captured in the temporal dynamics of $K_I$ and $K_A$ governed by the exchange rate $k$ as described by the MGE framework:

$$\ln E \approx -bD + \frac{1}{6}b^2 D^2 b_\mu^2 K_{int} + \frac{1}{6}b^2 D^2 h(k)[K_I^0 + h_\Delta^2(k)K_A^0] \qquad (35)$$

Note that if exchange is fast, and assuming no non-exchanging compartments, the notion of distinct compartments above becomes void, and the system reduces to a single unit with total kurtosis given by $K_{int}$.

The possible presence of some non-exchanging compartments in the system, which would give rise to non-vanishing long-time kurtosis sources, is accounted for by incorporating $K_I^\infty$ and $K_A^\infty$ into Eq. 35 to yield

$$\ln E \approx -bD + \frac{1}{6}b^2D^2h(k)[K_I^0 + h_\Delta^2(k)K_A^0] + \frac{1}{6}b^2D^2[K_I^\infty + b_\Delta^2 K_A^\infty + b_\mu^2 K_{int}] \quad (36)$$

We refer to the unified signal representation in Eq. 36 as µMGE. The intrinsic kurtosis captured by $K_{int}$ is insensitive to exchange.

The mechanism by which µMGE separates exchange from intrinsic kurtosis is illustrated in Fig. 2 (assuming zero anisotropy for simplicity). The SDE measurement gives the total kurtosis containing all contributions, including the intrinsic kurtosis and is given by

$$K_{SDE} = K_I^0 h_{SDE}(k, \Delta) + K_I^\infty + K_{int} \quad (37)$$

Further, µMGE assumes that, for the process giving rise to $K_{int}$, the long-mixing-time regime is attained already at short mixing times. As such, any DDE measurement reduces the impact of $K_{int}$ to half (because $b_\mu^2 = \frac{1}{2}$ for DDE). It also incurs an additional mixing-time-dependent reduction due to exchange:

$$K_{DDE} = K_I^0 h_{DDE}(k, \Delta, t_m) + K_I^\infty + \frac{1}{2}K_{int} \quad (38)$$

That this allows separation of the intrinsic kurtosis can be understood from a thought experiment. Assume we use multiple-mixing-time DDE measurements to fit $k$ and a dummy parameter that holds the sum of $K_I^\infty$ and $\frac{1}{2}K_{int}$. We can then use this to predict almost all of the kurtosis of an SDE measurement, according to

$$K_{SDE}^{pred} = K_I^0 h_{SDE}(k, \Delta) + \left(K_I^\infty + \frac{1}{2}K_{int}\right) \quad (39)$$

Subtracting this predicted value from the observed one yields the intrinsic kurtosis:

$$K_{SDE} - K_{SDE}^{pred} = \frac{1}{2}K_{int} \quad (40)$$

A connection between this picture and CTI can be made by noting that the CTI-estimated microscopic kurtosis would be given by

$$K_{SDE} - K_{DDE}(t_m \to \infty) = \frac{1}{2}K_I^0 h_{SDE}(k, \Delta) + \frac{1}{2}K_{int} = \frac{1}{2}K_\mu \quad (41)$$

Alternatively,

$$K_\mu = K_I^0 h_{SDE}(k, \Delta) + K_{int} \tag{42}$$

Thus, microscopic kurtosis contains contributions from intercompartmental exchange as well as intrinsic kurtosis. Note that Eq. 42 (and indeed any other equation involving the exchange-weighting function $h$) requires that $k$ is different from zero. In the limit $k \to 0$, $K_\mu$ approaches the sum $K_I^0 + K_{int}$.

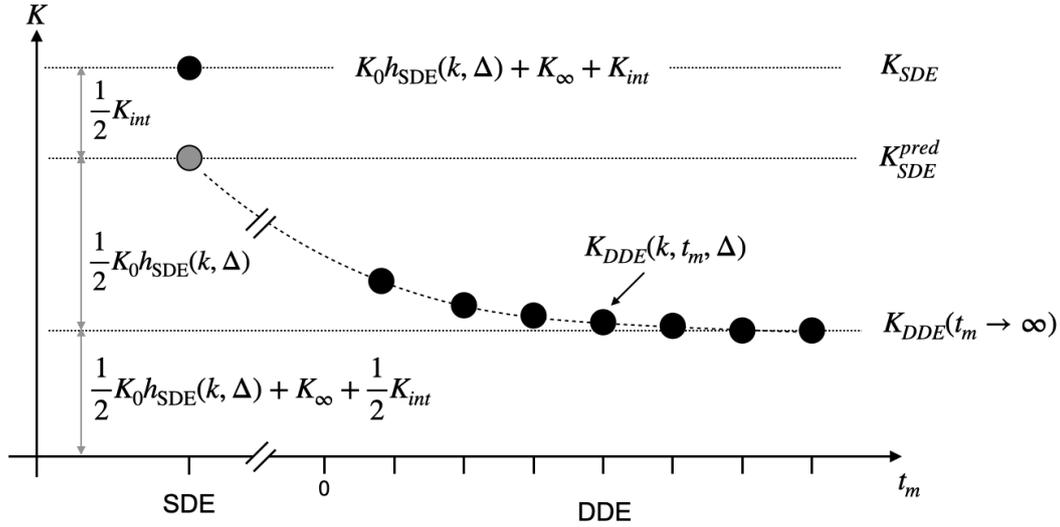

Figure 2: Principle of µMGE. The SDE measurement gives the total kurtosis including all contributions. DDE measurements at any mixing time reduce the kurtosis by half the intrinsic value plus exchange-driven attenuation. The DDE points allow estimation of $K_0$ and $k$ which in turn allow prediction of the SDE measurement under the assumption of zero intrinsic kurtosis. Any deviation between the predicted and observed SDE kurtoses manifests as intrinsic kurtosis.

The above discussion highlights that µMGE requires at least two different mixing times to facilitate the DDE-based prediction of the SDE kurtosis. With only one mixing time, the separation of $k$ from $K_{int}$ is impossible. It should also be noted that µMGE suffers from the same degeneracies as MGE at zero exchange. In addition, while MGE can be applied to experiments with arbitrary gradient waveforms, µMGE is restricted to DDE setups with fixed pulse durations and diffusion time in each block but variable mixing times, due to the assumptions in CTI and the definition of $b_\mu^2$.

# 3 Methods

*Protocol design*

Synthetic signals were generated using four different acquisition protocols. Three of these used gradient strengths typical of preclinical systems while the fourth was adapted to a typical clinical scanner. The first protocol was designed as described in [41]. The protocol comprised four sets of acquisitions with different combinations of $b_1$, $b_2$ and $\theta$ (Fig. 3A and B). Set 1 had $b_1 = 2.5$ ms/µm², $b_2 = 0$ and 45 rotations chosen to minimise electrostatic repulsion on a sphere. Set 2 had $b_1 = b_2 = 1.25$ ms/µm², $\theta = 0°$ and rotated as in Set 1. Set 3 had $b_1 = b_2 = 1.25$ ms/µm², $\theta = 90°$, rotated as in Set 1. The acquisition was repeated for three equidistant directions of the second gradient pair, yielding in total 135 rotations for Set 3. Set 4 had $b_1 = b_2 = 0.5$ ms/µm², $\theta = 0°$ and rotated as in Set 1. Pulse timing parameters were $\delta = 3.5$ ms and $\Delta = t_m = 12$ ms.

A second protocol was designed to test the fulfilment of the long-mixing-time regime required for CTI, as proposed by [41]. This protocol featured the same general timing settings as the first protocol, but only two gradient waveforms designed to yield parallel and anti-parallel DDE, according to $(b_1, b_2, \theta) = (1$ ms/µm², $1$ ms/µm², $0°)$ and $(b_1, b_2, \theta) = (1$ ms/µm², $1$ ms/µm², $180°)$.

A third protocol was created to assess the effects of a variable mixing time on $K_\mu$ from CTI and to evaluate MGE (Eq. 32) and µMGE (Eq. 36). This protocol was identical to the first one, except that it featured eleven mixing times for both parallel and orthogonal DDE (1, 4, 8, 12, 16, 20, 30, 50, 100, 200 and 300 ms) and six b-values for both SDE and DDE acquisitions (0.25, 0.5, 1, 1.5, 2 and 2.5 ms/µm²).

Finally, a fourth protocol was created with timing parameters $\delta/\Delta/t_m = 15.8/31.8/32.3$ ms obtained from [72] where it was implemented on a 3T clinical scanner. In the present work, this protocol used the same b-values and rotation scheme as the third protocol. To enable estimation of µMGE parameters with this protocol, a range of additional mixing times was added to yield $t_m = (1, 4, 8, 12, 16, 20, 32.3, 50$ and $100$ ms) for both parallel and orthogonal DDE.

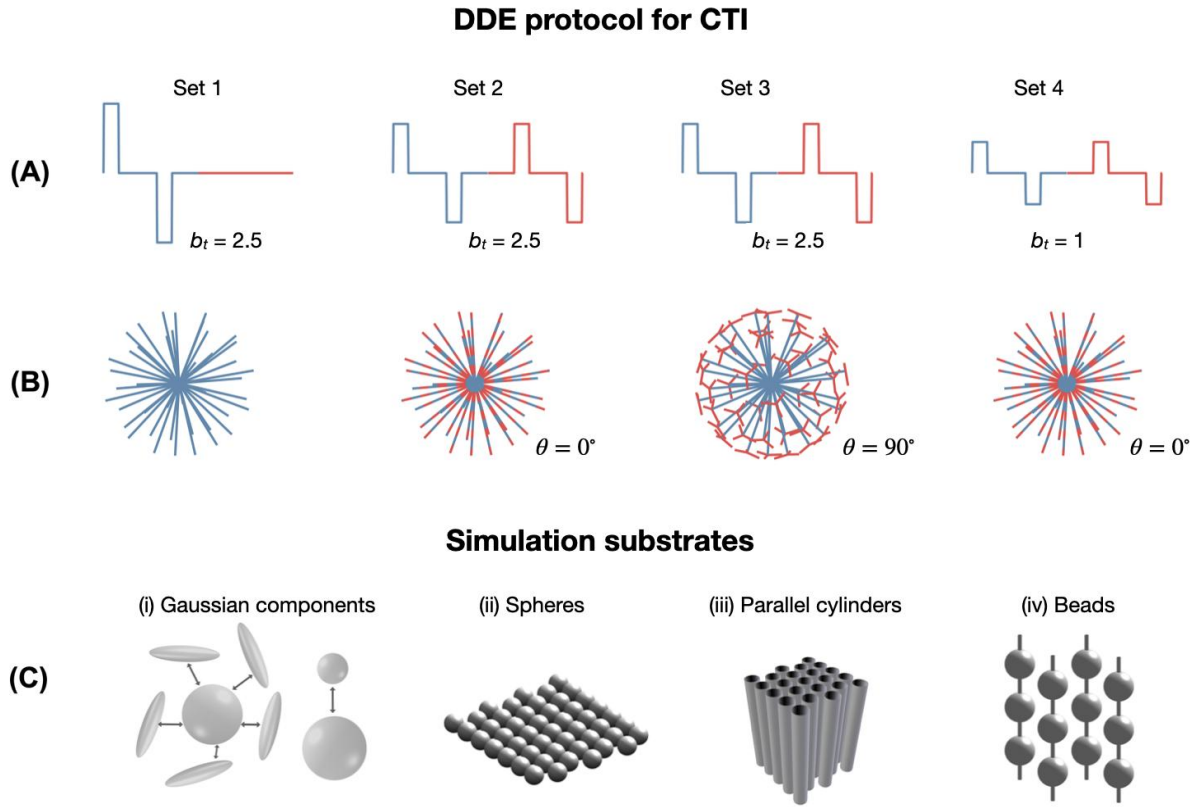

Figure 3: Protocols and substrates used in simulations. (A) shows four sets of DDE waveforms with different combinations of b-values $b_1$ and $b_2$, $b_t = b_1 + b_2$ denotes the total b-values and $\theta$ is the angle between the first and second gradient pairs. (B) shows the rotation schemes used in each acquisition set. Set 1 has $(b_1, b_2, \theta) = (2.5,,0,0°)$ rotated in 45 directions chosen to minimise electrostatic repulsion on a sphere. Set 2 has $(b_1, b_2, \theta) = (1.25,1.25,0°)$ rotated as in set 1. Set 3 has $(b_1, b_2, \theta) = (1.25,1.25,90°)$ rotated as in set 1 and repeated for three equidistant directions of the second gradient pair. Set 4 has $(b_1, b_2, \theta) = (0.5,0.5,0°)$ rotated as in set 1. Panel (C) shows the four simulation substrates used: sets of isotropic and anisotropic Gaussian components in exchange with each other, regularly packed spheres of diameter 6 μm, regularly packed parallel cylinders of diameter 1 μm and beading structures with maximum and minimum diameters of 6 μm and 1 μm.

*Microstructure simulations*

In order to investigate the effect of exchange on CTI-estimated microscopic kurtosis and to evaluate the μMGE approach, Monte Carlo simulations were performed in three different substrates: regularly packed spheres of diameter 6 μm (packing density = 50%), regularly packed cylinders of diameter 1 μm (packing density = 75%) and beading structures with minimum and maximum diameters of 1 and 6 μm (packing density = 40%), all in exchange with the extracellular space (Fig. 3C). Barrier permeability was varied to yield exchange rates between 0 and 200 s$^{-1}$. Other simulation parameters were: number of particles = $10^6$, temporal resolution = 0.5 μs, bulk diffusivity = 2 μm$^2$/ms and intra- and extracellular particle populations were initialised to maintain equal particle densities in each compartment. In each simulation scenario, signals were generated using the protocols described in the preceding section. The simulations were executed using an inhouse-written GPU-accelerated simulation framework available at https://github.com/arthur-chakwizira/Pasidi. For details, see the supplementary material of [51].

*Gaussian simulations*

While the microstructure simulations described above are more realistic than simulations of purely Gaussian diffusion, they lack a known ground truth intrinsic kurtosis and potentially feature restriction-induced diffusion time-dependence which is not accounted for by the MGE theories considered in this work. For this reason, Monte Carlo simulations were also performed in Gaussian components featuring either isotropic-to-isotropic exchange ("iso-iso Gaussian") or isotropic-to-anisotropic exchange ("iso-aniso Gaussian") (Fig. 3C). In both cases, the ground truth intrinsic kurtosis of individual components was considered to be zero. These simulations used the same number of particles and temporal resolution as the microstructure simulations but featured two pools with diffusion tensors given by $[D_{iso}, D_\Delta]$ = [2 μm$^2$/ms, 0] and [0.5 μm$^2$/ms, 0] for the "iso-iso" case and $[D_{iso}, D_\Delta]$ = [0.5 μm$^2$/ms, 1] and [1.5 μm$^2$/ms, 0] for the "iso-aniso" case. The anisotropic tensor was oriented along the x-direction. Pool fractions were set to 50 % in each case. The same simulation framework that was used for the microstructure simulations was also used for the Gaussian case. It is worth noting that the Gaussian simulations described above are captured by the SMEX/NEXI models presented in previous work [71,73].

An additional multi-Gaussian simulation was designed specifically to illustrate the mechanism of μMGE. This substrate featured two isotropic components with $[D_{iso}, D_\Delta]$ = [0.1 μm$^2$/ms, 0] and [0.5 μm$^2$/ms, 0] with rapid exchange between them ($k$ = 100 s$^{-1}$).

These two components were then in slow exchange ($k = 0 - 10$ s$^{-1}$) with a third component which had $[D_{iso}, D_\Delta] = [1$ μm²/ms, $0]$. The fractions of the three components were set to 0.25, 0.25 and 0.5, respectively. The idea was that the rapidly exchanging components would give rise to a combined local unit that effectively exhibited non-zero intrinsic kurtosis. Due to the different timescales of the two processes, the intrinsic kurtosis should in principle be separable from the much slower intercompartmental exchange using the μMGE approach.

*Error propagation*

The μMGE theory contains many free parameters. It is thus relevant to determine whether all parameters could be estimated from noisy signals. This was achieved by studying the dependence of the coefficients of $K_I^0$, $K_A^0$ and $K_{int}$ in Eq. 36 on mixing time and exchange rate. The rationale was that any correlation between these coefficients would result in an underdetermined equation system which makes it challenging to disentangle the parameters they encode for. For this part of the study, the exchange rate was varied from 0 to 200 s$^{-1}$. The second part of the feasibility study aimed at testing the ability to solve the inverse problem in Eq. 36 and involved generating signals with the μMGE signal representation at different exchange rates using protocol 3 with mixing times up to 100 ms, corrupting the signals with Rice-distributed noise at SNR = 200 and fitting μMGE back to those signals. The exchange rate was varied between 0.1 and 200 s$^{-1}$, $K_{int}$ was set to 0.5, $K_I^0$ and $K_A^0$ were both set to 1, and $K_I^\infty$ and $K_A^\infty$ were both set to 0.5. Performance of μMGE was evaluated using the accuracy and precision of parameter estimates.

*Data analysis*

Signals simulated using the protocols and substrates in Fig. 3 were powder-averaged and CTI parameter estimates were obtained by fitting Eq. 10. Fulfilment of the long-mixing-time condition was checked by computing the difference between the log of powder-averaged signals obtained with the parallel and antiparallel DDE protocols [41]. 1D-MGE parameter estimates were obtained by fitting Eq. 13 to a subset of the powder-averaged data with $\theta = 0°$. MGE (Eq. 32) was evaluated using signals generated with protocol 3 (featuring multiple mixing times and b-values) in substrates of spheres, iso-iso Gaussian and iso-aniso Gaussian. The signals were corrupted with Rice-distributed noise at a generous SNR of 200 prior to fitting. Numerical evaluation of µMGE was done by fitting Eq. 36 to signals generated using the protocols 3 and 4 in all substrates shown in Fig. 3 using exchange rates between 5 and 200 s$^{-1}$. In each case, µMGE kurtosis estimates were compared with the CTI kurtosis estimates under the same simulation settings.

Throughout this study, fitting was performed using the method *lsqnonlin* in MATLAB (The MathWorks, Natick, MA, R2022a). The fits were initialised at ten randomly chosen starting conditions between the upper and lower bounds, and the fit with the least residuals was selected as the solution. All fitting code and the simulated signals can be found in the repository at https://github.com/arthur-chakwizira/cti-mge.

# 4 Results

Figure 4 shows powder-averaged CTI signals obtained using the protocols shown in Fig. 3 in four simulation substrates. Corresponding estimates of $K_T$, $K_I$, $K_A$ and $K_\mu$ are shown alongside the signals. Results are shown in the absence of exchange (column A) and in the presence of exchange at a rate of 50 s$^{-1}$ (column B). In the absence of exchange, there is a small positive $K_\mu$ in the substrates of spheres, cylinders and beads. The Gaussian case gives zero $K_\mu$, as expected. In the presence of exchange, there is a large positive $K_\mu$ in all substrates. The observed total, isotropic and anisotropic kurtosis decrease with the introduction of exchange. In a substrate of spheres with only intracellular particles, the microscopic kurtosis is negative in alignment with expectations (c.f Fig. A1 of the supplementary material). In the substrate mimicking beaded axons, a large positive $K_\mu$ was found (Fig. A1).

Figure 5 shows how CTI kurtosis estimates respond to variations in the underlying exchange rate in for the substrates of spheres of diameter 6 µm and isotropic exchanging Gaussian components. The exchange-driven decline in isotropic kurtosis is captured by $K_\mu$, which grows with the underlying exchange rate up to a peak after which it decreases, in accordance with the prediction of Eq. 19. At non-zero exchange, $K_\mu$ increases with the mixing time (indicating violation of the long mixing time condition of CTI) and plateaus at long mixing times, also in alignment with the theory (compare Fig. 1 and Eq. 19). Figure 5 also shows kurtosis and exchange estimates obtained with 1D-MGE (Eq. 13), of which the kurtosis is independent of the mixing time and the exchange estimates correlate with the simulated values. These results indicate that the substrate of exchanging spheres is well-approximated by the multi-Gaussian assumption. Similar trends were observed with the clinical protocol (c.f. Fig A2 of the supplementary).

The parallel-antiparallel log signal difference was 0.0009, -0.0005 and 0.0012 for the substrates of spheres, beads and cylinders, at a mixing time of 12 ms, which are all close to zero, suggesting that the long-mixing time regime was attained. However, according to Fig. 5, $K_\mu$ is independent of the mixing time only for mixing times above 100 ms, which suggests that the parallel-antiparallel signal difference reported above may not be a sufficient test for the fulfilment of the long mixing time regime of CTI.

Figure 6 shows MGE parameter estimates in exchanging Gaussian components that are either isotropic or anisotropic. Kurtosis estimates are largely independent of the underlying exchange rate and largely agree with the ground truth (although some bias and low precision are evident at higher exchange rates). Exchange rate estimates obtained with MGE correlate well with the ground truth but with a slight bias, which is attributable to the influence of higher order terms in the cumulant expansion [51]. Similar results were obtained in the substrate of spheres and are shown in Fig. A3 of the supplementary material. Note that the long-time kurtosis parameters display a more pronounced exchange-rate dependence with decreasing exchange rate, illustrating the innate degeneracy of MGE (Eq. 32) at slow exchange. The behaviour of the estimates as the exchange rate approaches zero is shown in Fig. A4 of the supplementary material.

Figure 7 shows the results of the μMGE feasibility study. The coefficients of the different model parameters in Eq. 36 are shown in panel A and indicate that at zero exchange it is not possible to disentangle the isotropic and anisotropic kurtoses from their long-time variants. At intermediate exchange rates, however, the coefficients are independent for all parameters of the representation. At very high exchange rates, the mixing time dependence is again lost for all coefficients and the remaining contrast is only between SDE and DDE. This makes it difficult to disentangle intrinsic kurtosis from very fast exchange. Panel B of Fig. 7 shows the bias and precision (standard deviation) in parameter estimates obtained by fitting Eq. 36 to noisy signals generated using the same equation. In alignment with panel A, µMGE shows poor performance (large bias and low precision) at very slow and very fast exchange, but acceptable performance for the intermediate regime. These simulations were performed using the same model in the forward and inverse problem and show that the inverse problem can be solved with precision only in the intermediate regime. However, they do not report on the applicability of the model to any specific type of microstructure.

CTI and µMGE parameter estimates in the multi-Gaussian substrates that was designed to exhibit non-zero intrinsic kurtosis according to the definitions in this work are shown in Figure 8 (for protocol 1). When the intercompartmental exchange rate ($k_{true}$) is set to zero, both CTI and µMGE measure a non-zero intra-compartmental kurtosis sensitive to the rapid exchange between two of the Gaussian components. As $k_{true}$ increases, CTI $K_\mu$ also increases, illustrating that microscopic kurtosis comprises both the intrinsic kurtosis and intercompartmental exchange. On the contrary, µMGE $K_{int}$ maintains its value at $k_{true} = 0$, with variations in $k_{true}$ being captured by the exchange estimate $k$. This illustrates the ability of µMGE to separate intercompartmental exchange from intrinsic kurtosis. Note that, for µMGE, $K_I$ and $K_A$ denote the sums $(K_I^0 + K_I^\infty)$ and $(K_A^0 + K_A^\infty)$, respectively.

Figure 9 shows CTI parameter estimates in four substrates: permeable spheres, permeable beads and exchanging Gaussian components with either isotropic or anisotropic diffusion. The figure corresponds to signals obtained using protocol 1 (preclinical) and protocol 4 (clinical). In all substrates, kurtosis estimates from CTI vary with the exchange rate and also depend on the protocol. In particular, $K_\mu$ exhibits the typical non-monotonic dependence on exchange rate illustrated in Fig. 1. Corresponding estimates for µMGE are presented in Fig. 10 where the kurtosis estimates are independent of the exchange rate for the preclinical protocol but become a function of the exchange rate above 50 s$^{-1}$ with the clinical protocol. The failure of µMGE at faster exchange is due to the loss of the mixing-time-dependence as illustrated in Fig. 7, which happens at lower exchange rates when the diffusion time is long. This is further confirmed by analysing the same data using only MGE (Fig. A6), which largely removes the dependence of the kurtosis estimates on the underlying exchange rate. Exchange estimates from µMGE generally correlate well with the ground truth. Similar trends were observed in the substrate of cylinders, and the result is presented in Fig. A5 of the supplementary material.

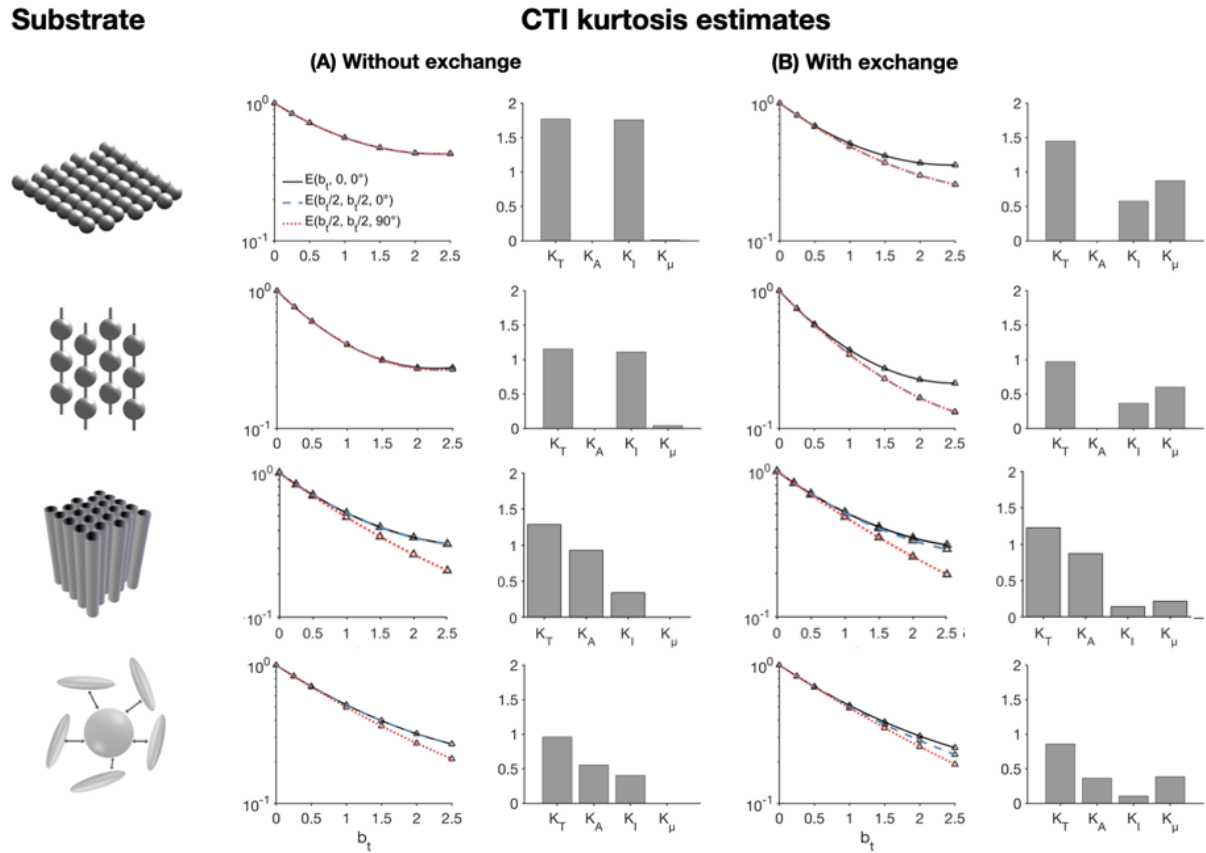

Figure 4: CTI simulated signals and parameter estimates. The first column shows the simulation substrates used for generation of each of results in (A) and (B). (A) shows powder-averaged signals obtained with the CTI protocol in Fig. 3 (protocol 1) for the case of zero exchange. Corresponding CTI estimates of total, isotropic, anisotropic and microscopic kurtosis are shown alongside the signals. The same results are presented in (B) but in the presence of exchange at a rate of 50 s$^{-1}$. Microscopic kurtosis is small in all substrates in (A) but increases notably when exchange is introduced in (B).

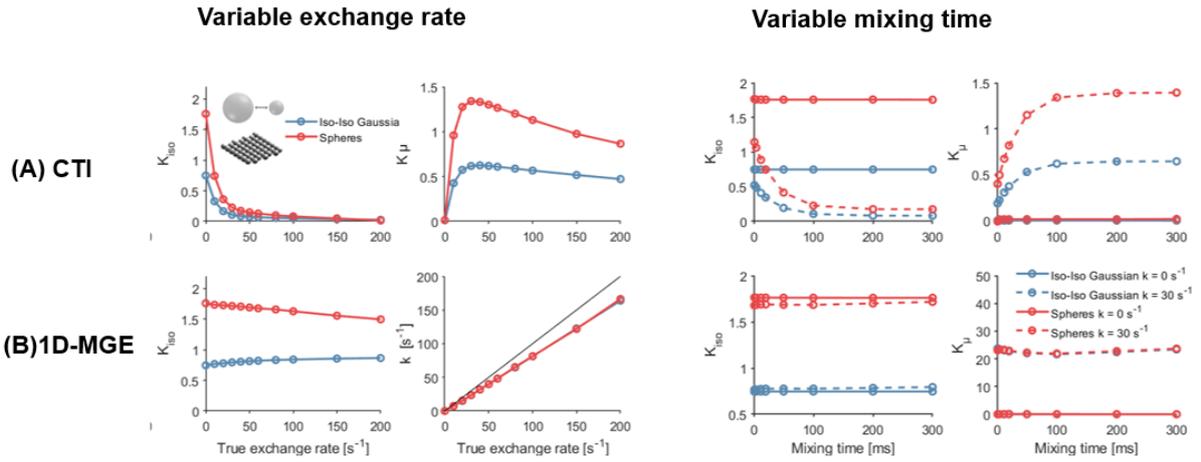

Figure 5: Variation of CTI and 1D-MGE parameter estimates with exchange rate and mixing time. Results are shown in substrates of regular spheres in exchange with the extracellular space and isotropic Gaussian components in exchange. In both substrates, CTI-estimated isotropic kurtosis decreases with the exchange rate, while microscopic kurtosis increases up to a peak after which it decreases. CTI-estimated microscopic kurtosis also increases with the mixing time up to a plateau when exchange is non-zero, in line with the prediction of Eq.19. Note that the results in the top row of (A) were generated using protocol 3 with a mixing time of 100 ms. Exchange estimates obtained with 1D-MGE show an independence on mixing time (B) and correlate with the ground truth.

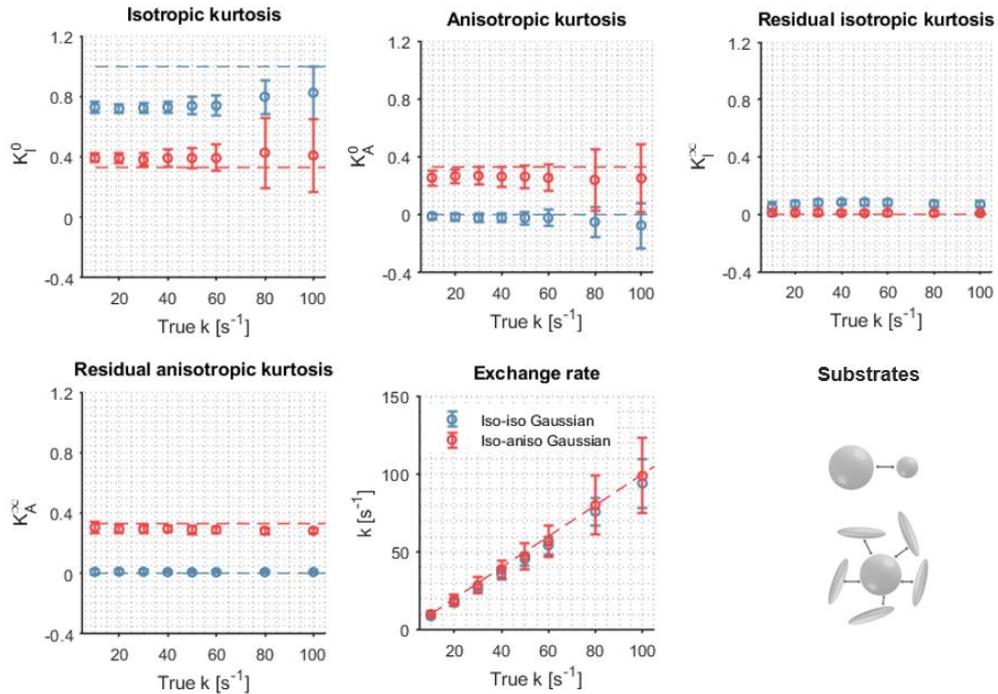

Figure 6: Evaluation of MGE in different substrates: isotropic Gaussian components in exchange (Iso-Iso Gaussian) and anisotropic Gaussian components in exchange with an isotropic component (Iso-Aniso Gaussian). Estimates were obtained from signals generated using protocol 3 at an SNR of 200. The error bars represent one standard deviation. The kurtosis estimates show – as expected – a weak to no dependence on the true exchange rate. MGE exchange estimates in all three substrates agree well with the ground truth. However, MGE shows a notable decline in precision at high exchange rates.

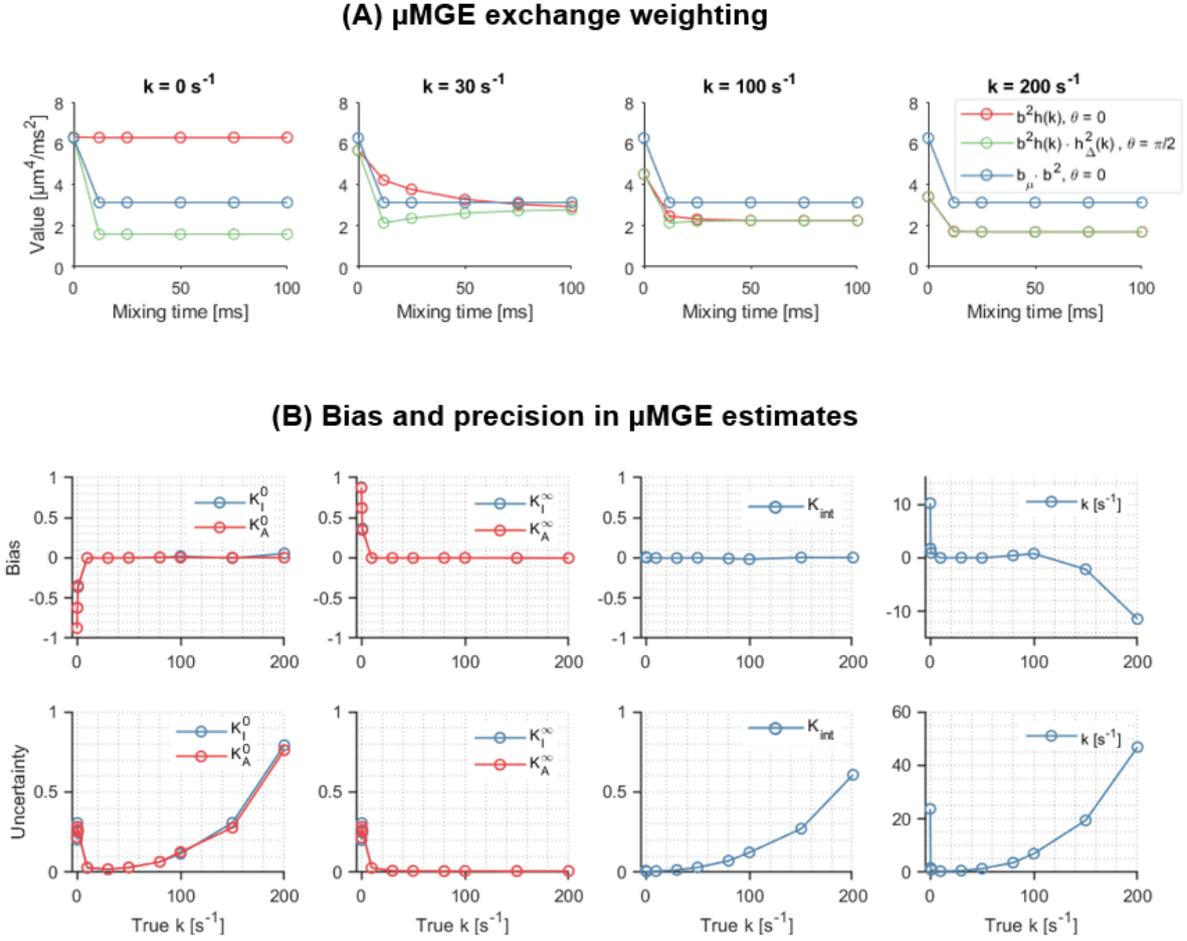

Figure 7: Study of the invertibility of the μMGE representation. (A) shows the dependence of the coefficients of $K_I^0$, $K_A^0$, and $K_{int}$ in Eq. 36 on mixing time for different exchange rates. The lack of dependence on mixing time of the coefficients of $K_I^0$ and $K_A^0$, at zero exchange rate means that these parameters cannot be disentangled from their long-time variants $K_I^\infty$ and $K_A^\infty$. At intermediate exchange rates, all parameters of Eq. 36 can be estimated from the signals. At very fast exchange (k = 200 s⁻¹), the dependence on mixing time is again lost, making it difficult to disentangle the exchange rate from intrinsic kurtosis. Panel B shows bias and precision (1 std. dev.) in parameter estimates obtained by generating signals using the μMGE representation (Eq. 36), corrupting the signals with Rice-distributed noise at SNR = 200 and fitting the same equation to those signals. The figure shows high bias and low precision at very slow and fast exchange rates, indicating inability to invert the μMGE representation in these extremes. The representation is, however, invertible at intermediate exchange rates.

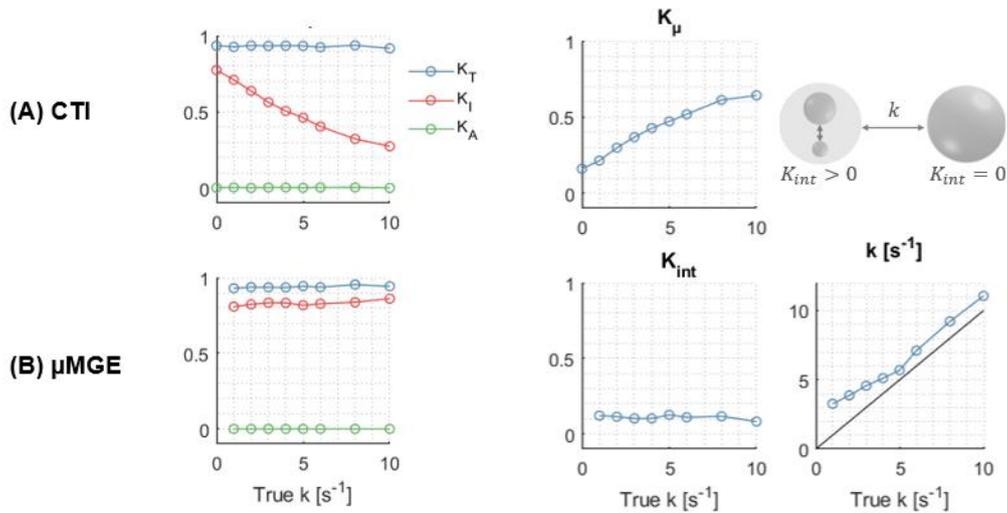

Figure 8: Comparison of CTI and μMGE in a substrate designed to exhibit non-zero intrinsic kurtosis. Increasing the intercompartmental exchange rate causes a decline in CTI-estimated isotropic kurtosis, which is captured by an increase in the microscopic kurtosis. μMGE shows an exchange-independent, non-zero intrinsic kurtosis and captures the intercompartmental exchange via the exchange rate $k$.

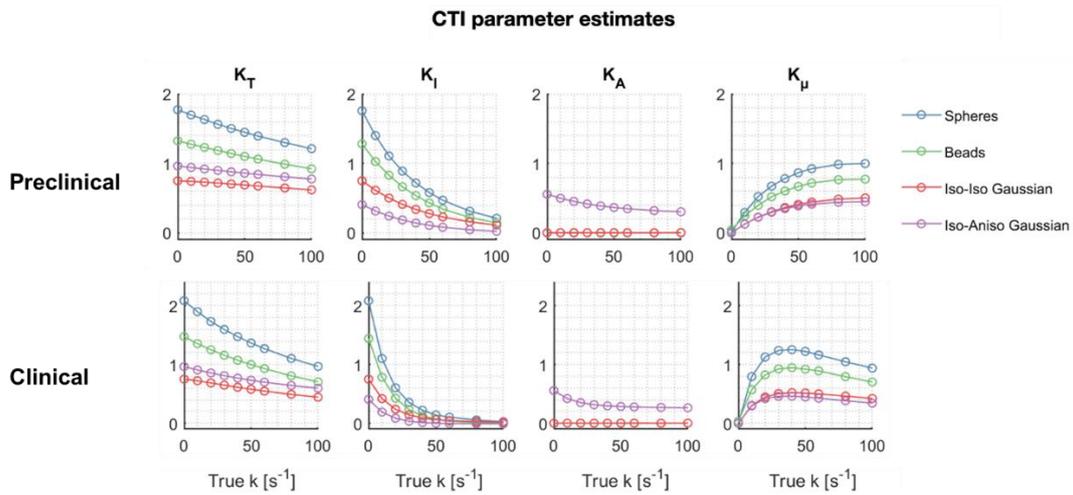

Figure 9: Evaluation of CTI in different substrates and with different protocols. Results are shown for noiseless signals generated in spheres in exchange with the extracellular space, beads in exchange with the extracellular space, exchanging isotropic Gaussian components and anisotropic Gaussian components in exchange with isotropic. All CTI kurtosis estimates vary with the exchange rate for both the preclinical and clinical protocols.

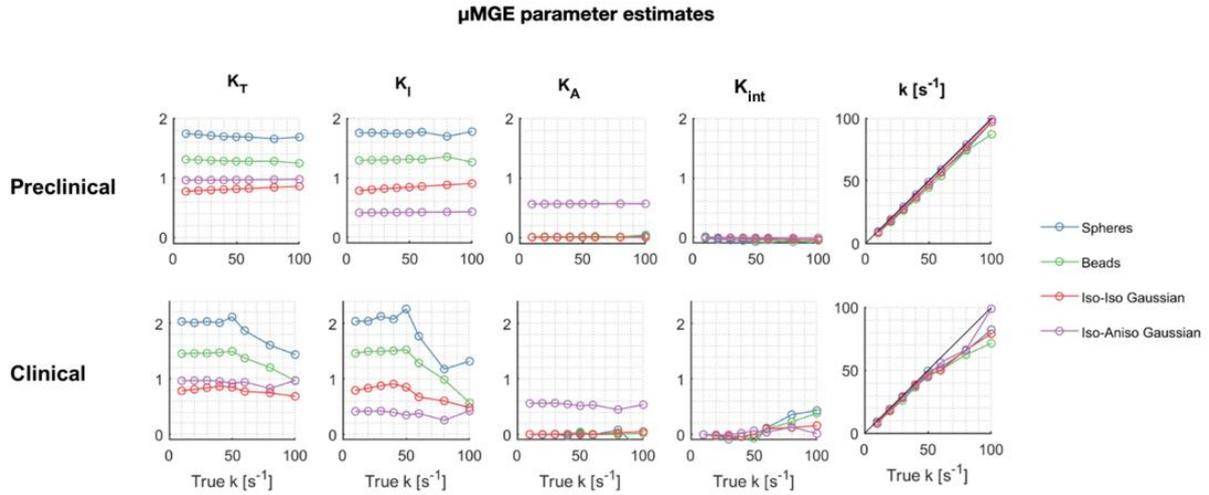

Figure 10: Evaluation of μMGE in different substrates and with different protocols. Results are shown for noiseless signals generated in spheres in exchange with the extracellular space, beads in exchange with the extracellular space, exchanging isotropic Gaussian components and anisotropic Gaussian components in exchange with isotropic. Kurtosis estimates for the preclinical protocol are independent of the exchange rate over the range shown. With the clinical protocol, kurtosis estimates begin to depend on the exchange rate above 50 s$^{-1}$.

# 5 Discussion

This study has shown using both theory and simulations (Eq. 19, 21, Fig. 4-5) that exchange is a source of the microscopic kurtosis estimated with CTI. In the presence of exchange, $K_\mu$ will thus depend on the mixing time when the long-mixing time condition is not satisfied and in the long-mixing time regime $K_\mu$ will depend on the diffusion time (Fig. 5, 8). In comparison, the exchange rates as estimated by 1D-MGE are largely invariant to both the mixing time and the diffusion time (Fig. 5, Fig. 10). However, the 1D-MGE approach is not a viable approach in systems with anisotropy, and the exchange rate it estimates can be biased by kurtosis sources not included in the approach. To address this, we extended 1D-MGE in two steps. First, we developed the MGE theory which accounts for anisotropy and residual kurtosis and evaluated it with simulations (Fig. 6). Second, we introduced intrinsic kurtosis into MGE and formed the μMGE approach, which allowed us to disentangle intercompartmental exchange from intrinsic kurtosis (Fig. 7, 8, 10). While this approach could in principle be applied in future work to simultaneously map intrinsic kurtosis and exchange separately, it is a demanding experimental approach. This begs the question of whether new studies based on data acquired with DDE should apply CTI, 1D-MGE, MGE, or µMGE in their data analysis. The answer depends on the situation, and we will provide perspectives to inform the decision.

Exchange estimation with diffusion MRI has garnered increasing interest in recent years. Several studies have applied FEXI to map exchange in the healthy human brain [46,50,74], brain tumors [50], breast cancer [75] and in animal models on preclinical scanners [76,77]. Recent work has also shown sensitivity to blood-brain-barrier exchange [74,78,79]. Other studies have applied diffusion-exchange spectroscopy (DEXSY) to measure relatively high exchange rates in the ex-vivo mouse brain [80,81]. A new class of models based on a combination of the standard model and the Kärger model has been recently developed and applied to map exchange in the rat brain using SDE with variable diffusion times [71,73]. More recent work has demonstrated the utility of free gradient waveforms for mapping exchange in the human brain unconfounded by restricted diffusion [51,82]. The general consensus regarding exchange in the brain is that it is small and potentially negligible between the intra- and extra-axonal spaces in healthy myelinated white matter at clinically-accessible diffusion times and this finding is explained by the presence and low permeability of myelin sheaths [83–87]. In ischemic brain tissue, relatively short exchange times between approximately 50 and 500 ms have been reported [88,89]. Gray matter is more complicated. The powder-averaged signal at strong diffusion weighting deviates from the typical inverse-square-root dependence on the b-value characteristic of impermeable sticks which indicates non-negligible exchange between neurites and the extracellular space or neurites and somas [73,87,90]. Exchange estimates in gray matter are, however, highly variable, with literature values ranging from 4 ms to hundreds of milliseconds [73,87]. The theory (Eq. 19) predicts that non-zero intrinsic kurtosis may manifest as exchange even in the absence of exchange through permeable cell membranes, suggesting that the shorter exchange time estimates reported above were possibly influenced by intrinsic kurtosis.

CTI is a novel technique that leverages dedicated DDE experiments to probe the displacement correlation tensor containing information about isotropic and anisotropic kurtosis [40,41]. A subtraction of the isotropic and anisotropic components from the total kurtosis provides the microscopic kurtosis—a kurtosis component assumed to be zero in previous approaches based on the multi-Gaussian diffusion assumption [30,31,40,41].

Simulations previously used to validate CTI were mostly based on an introduction of microscopic kurtosis via an analytical signal decay [41] and are thus difficult to compare to the present work that performs Monte Carlo simulations designed to mimic a realistic dMRI experiment of tissue with water in both intra- and extracellular spaces. It is worth noting, however, that the supplementary material of [41] reports results of Monte Carlo simulations inside spheres where the estimated negative $K_\mu$ (approximately –0.4) is in good agreement with our estimates (Fig. A1). More recent work investigates CTI using Monte Carlo simulations inside beading structures with varying beading amplitudes but without extracellular water [91] where large positive $K_\mu$ (approximately 5) was observed in the substrate with the highest degree of beading. The findings of the present work (Fig. A1) are consistent with these previous results. Moreover, Alves et al. [91] report separately on $K_\mu$ estimates from the intracellular and extracellular spaces of the beading structures, where the latter shows $K_\mu$ close to zero. Including both intracellular and extracellular signals in the estimation of $K_\mu$ resulted in negligible $K_\mu$ given a typical bulk diffusivity of 2 μm²/ms—also in agreement with the present work. The vanishing of $K_\mu$ upon inclusion of extracellular spins is due to the relatively low packing density of the beads, which gives predominantly free diffusion with zero intrinsic kurtosis. Microscopic kurtosis is given by the average of the intrinsic kurtoses scaled by the inverse square of the mean diffusivity, and adding more free water increases the mean diffusivity without contributing much of an intrinsic kurtosis, which results in a reduced microscopic kurtosis.

Since its conception, CTI has been applied in various conditions. For example, $K_\mu$ was mapped in healthy volunteers where it was larger in cortical grey matter than in white matter [72]. In a mouse model of stroke lesions, a large increase in $K_\mu$ (between 50 and 100%) was observed in the ischemic regions and the authors attribute this increase to higher cell cross-sectional variance or restricted diffusion [91]. In summary, values of $K_\mu$ of up to 0.5 were reported in the healthy human brain and up to about 1 in stroke lesions, both of which are higher than what the present study detects in simulations featuring both intra- and extracellular signals. On this note, we would like to highlight that—as shown throughout this work (Figs. 4,5,8)—exchange is a source of $K_\mu$. The observations of elevated $K_\mu$ in grey matter and stroke lesions are *consistent* with previous reports of fast exchange in these tissues [82,88,89] and thus we find it likely that exchange is a more dominant contribution to the reported values of $K_\mu$ than intrinsic kurtosis. To empirically test this hypothesis, a protocol featuring variable mixing times for parallel DDE acquisitions can be used. Exchange would cause signal dependence on mixing time while intrinsic kurtosis would not.

At this point, it is worth clarifying how intrinsic kurtosis is defined and interpreted by the different methods. In a system of non-exchanging Gaussian components, intrinsic kurtosis is zero (the first term of Eq. 5 is zero). In MGE, upon the introduction of exchange into such a system, intrinsic kurtosis remains zero but the intercompartmental kurtosis undergoes a temporal decline driven by the exchange [47,51]. In CTI, the introduction of exchange into the multi-Gaussian system effectively creates a new "compartment" that encompasses all the exchanging components and that exhibits intra-compartmental heterogeneity that manifests as microscopic kurtosis [52]. This microscopic kurtosis is exchange-driven and thus depends on the underlying exchange rate, as has been shown in this work. Note that with this view, the mixing time must be much longer than the exchange time for the long mixing time conditions to apply. However, even in this limit, the exchange-driven microscopic kurtosis depends on the diffusion time (Eq. 21). Furthermore, this limit is not attainable from an SNR perspective if the exchange time is on the same order as the relaxation time that governs the signal loss during the mixing time. In µMGE, we attempt to disentangle the two contributions to microscopic kurtosis: intercompartmental exchange and intrinsic kurtosis (Fig. 8). While intercompartmental exchange is characterised by a signal dependence on the mixing time, intrinsic kurtosis should give no such dependence. Known sources of intrinsic kurtosis are restricted diffusion which gives a negative contribution and cross-sectional variation which gives a positive contribution [41,91]. The latter source presents an ambiguity because diffusion between domains of different diffusion properties within the same compartment carries the same signature as permeative exchange [68,92,93]. In this work, we defined a system with Gaussian diffusion and fast exchange between two components, which were then in slow exchange with the third. The difference in diffusivities between the components in fast exchange was considered as giving rise to the intrinsic kurtosis. This resembles two-compartment exchange, where the two pools in fast exchange would correspond to diffusive exchange between water close to and far away from the membrane, while the slow exchange would correspond to membrane permeation. However, such a system has additional degrees of freedom such as extracellular time dependence, and was thus unsuitable for our early investigation of the potential capacities of μMGE. Future work will be directed towards understanding how μMGE responds to geometrical configurations and permeabilities of cellular membranes.

If both intercompartmental exchange and intrinsic kurtosis influence the signal, a method to unambiguously estimate both effects would be valuable. This work presents a first step towards that goal (Eq. 36) by unifying the MGE theory with the concept of intrinsic kurtosis from CTI into the μMGE model. Simulations showed that the signal representation is invertible at intermediate exchange rates (Fig. 7) and that it disentangles intrinsic kurtosis from exchange in a variety of simulation substrates (Fig. 8-10). In the ambiguous case of diffusive and intercompartmental exchange highlighted above, μMGE can separate the two processes provided the diffusive exchange occurs on a much shorter timescale than the intercompartmental exchange (Fig. 8). The μMGE approach may thus provide a good candidate for an analysis framework in cases where both exchange and intrinsic kurtosis are expected to be relevant, such as in grey matter [72,82]. However, as highlighted earlier, the theory has an innate degeneracy at very low exchange rates (Fig. A4). The minimum exchange rate that can be estimated is ultimately determined by the longest mixing times employed, which in turn is ultimately limited by the relaxation-induced signal loss taking place during the mixing time.

Apart from analysing the interplay between exchange and microscopic kurtosis, we developed new theory to account for effects of anisotropy in exchange estimation. The directional dependence of estimated apparent exchange rates (AXRs) in anisotropic systems is known. Sønderby et al. [76] applied FEXI in multiple directions in both a yeast phantom and perfusion-fixated monkey brain and found that AXR was rotationally invariant in the yeast phantom but direction-dependent in anisotropic regions of the monkey brain. A theoretical study by Lasic et al. [94] revealed that AXR becomes anisotropic even in environments with a single exchange rate when there are more than two orientationally dispersed components. Li et al. [95] applied FEXI in multiple directions in human white matter and found that the AXR perpendicular to the fiber orientation was significantly larger than the AXR parallel to it. A simulation study by Ludwig et al. [96] showed that at least 30 gradient directions were required to reliably estimate AXR in the presence of orientation dispersion. We have presented an exchange theory that explicitly takes anisotropy into account, by uniting concepts from the previous 1D-MGE theory [47] and b-tensor encoding [30,31,64]. Numerical simulations in both isotropic and anisotropic substrates (Fig. 6) indicate that the new MGE theory correctly captures the exchange rate even in the presence of anisotropy, with the caveat that effects higher-order terms may cause a bias in kurtosis estimates. Another noteworthy aspect of the theory is the incorporation of residual time-independent kurtosis that may result from powder averaging or the presence of non-exchanging compartments in a voxel. This kurtosis component has been described in previous work [19,23,77] and the present work has illustrated its relevance with numerical simulations (Fig. 6). It should be reiterated, however, that the MGE theory is degenerate at low exchange rates and suffers from bias and low precision at high exchange rates (Fig. 6).

In conclusion, this work has demonstrated that microscopic kurtosis ($K_\mu$) estimated by CTI has contributions from both intrinsic kurtosis and intercompartmental exchange. We have presented and numerically evaluated new exchange theory (µMGE) accounting for anisotropy and enabling disentanglement of the two sources of microscopic kurtosis (exchange and intrinsic kurtosis). Our findings suggest that intrinsic kurtosis and exchange can be regarded as two separate phenomena that are distinguishable with appropriate modelling and experimental design.

# Data availability

All fitting code and the simulated signals can be found in the repository at https://github.com/arthur-chakwizira/cti-mge.

# 7 Appendix

## 7.1 On the relation between MGE and filter-exchange imaging

A relation between MGE and FEXI can be obtained by deriving the dependence of the ADC on the mixing time, defined according to

$$D'(t_m) = -\frac{1}{b_d}\left[\ln E_{f+d} - \ln E_f\right] \quad (A.1)$$

where $E_{f+d}$ and $E_d$ are the diffusion-weighted signals from acquisitions with the filtering and detection block active and detection only, respectively. These signals can be predicted using the MGE framework. We assume DDE acquisitions with short pulses which allows the exchange-sensitised square of the b-tensor to be written

$$\mathbf{B}^{\otimes 2}(k) = 2\int_0^T \mathbb{Q}_4(t)\exp(-kt)\,\mathrm{d}t = \mathbb{Q}_4^0\,\Delta h_0(k,\Delta) + 2\mathbb{Q}_4^1\,\Delta h_1(k,\Delta,t_m) \quad (A.2)$$

where $\mathbb{Q}_4^0 = \mathbb{Q}_4(0)$ and $\mathbb{Q}_4^1 = \mathbb{Q}_4(t_m + \Delta)$, $h_0(k,\Delta) = \frac{\Delta k + e^{-\Delta k} - 1}{\Delta^2 k^2}$ and

$$h_1(k,\Delta,t_m) = \left[\frac{e^{-t_m k} - e^{-k(\Delta+t_m)} - \Delta k e^{-k(\Delta+t_m)}}{\Delta^2 k^2}\right] + \left[\frac{e^{-kT} + kTe^{-k(\Delta+t_m)} - e^{-k(\Delta+t_m)} - k\Delta e^{-k(\Delta+t_m)} - kt_m e^{-k(\Delta+t_m)}}{\Delta^2 k^2}\right].$$

Assuming $k\Delta \ll 1$ gives $h_0(k,\Delta) \approx 1/2$ and $h_1(k,\Delta,t_m) \approx e^{-kt_m}$. $\mathbf{B}^{\otimes 2}(k)$ can then be written

$$\mathbf{B}^{\otimes 2}(k) = \mathbf{B}_f^{\otimes 2} + \mathbf{B}_d^{\otimes 2} + 2\mathbf{B}_f \otimes \mathbf{B}_d\, e^{-kt_m} \quad (A.3)$$

The isotropic and anisotropic projections of $\mathbf{B}^{\otimes 2}(k)$ can then be expressed

$$b_I^2(k) = (b_f^2 + b_d^2) + 2b_f b_d e^{-kt_m} \quad (A.4)$$

$$b_A^2(k) = (b_f^2 + b_d^2) + 2b'_\Delta b_f b_d e^{-kt_m} \quad (A.5)$$

where $b'_\Delta = 1$ for SDE and parallel DDE and $b'_\Delta = -\frac{1}{2}$ for orthogonal DDE. For simplicity, we re-write the MGE representation as:

$$\ln E \approx -bD + \frac{1}{2}\left[b_I^2(k)V_I + b_A^2(k)V_A\right] \quad (A.6)$$

Now the desired signal difference is given by

$$D'(t_m) = \langle D \rangle - \frac{1}{2} b_d [V_I + V_A] - \frac{1}{2} b_f e^{-kt_m}[V_I + b'_\Delta V_A] \quad (A.7)$$

At long mixing times, $t_m \to \infty$, the diffusivity approaches its equilibrium value

$$D^{eq} = \langle D \rangle - \frac{1}{2} b_d [V_I + V_A] \approx \langle D \rangle \quad (A.8)$$

where the approximation assumes $b_d$ is low. Defining the filter efficiency as

$$\sigma = 2b_f \frac{(V_I + b'_\Delta V_A)}{\langle D \rangle} \quad (A.9)$$

gives us the well-known FEXI signal equation:

$$D'(t_m) = D^{eq}[1 - \sigma e^{-kt_m}]. \quad (A.10)$$

Note that Eq. A.9 implies that the filter efficiency for orthogonal DDE can be negative if $V_A > 2 \cdot V_I$.

## 7.2 Supplementary figures

This section contains CTI kurtosis estimates in substrates of spheres and beads, where the intracellular and extracellular signals were analysed separately. MGE kurtosis and exchange estimates in multi-Gaussian components at slow exchange and in spheres that feature non-Gaussian diffusion with exchange are also presented herein. Finally, CTI and μMGE are compared in the substrate of parallel cylinders.

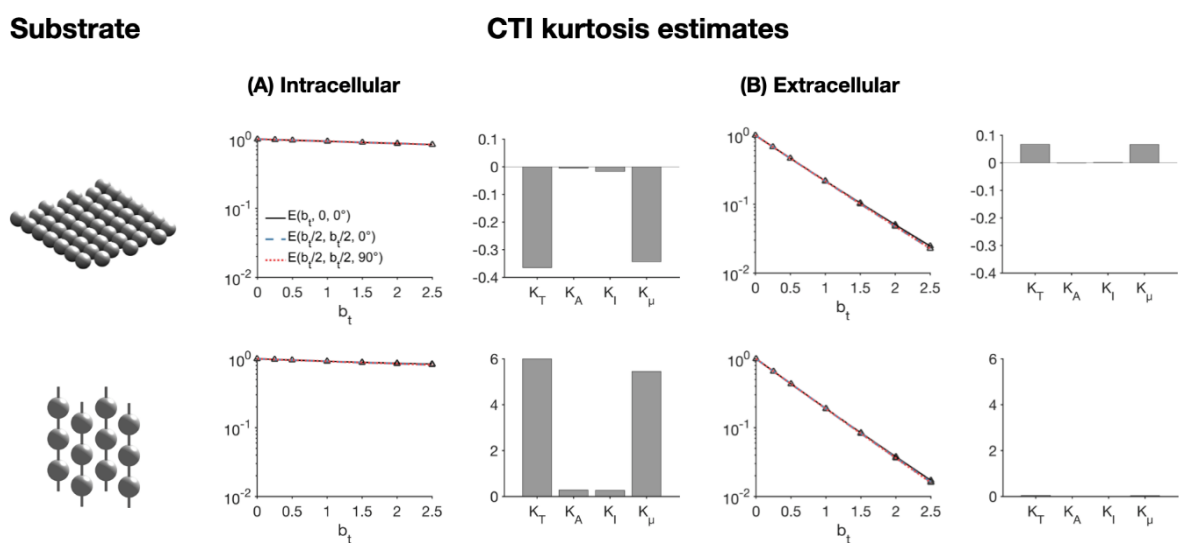

Figure A1: CTI parameter estimates in the substrates of 6 μm-diameter spheres and beads. The estimates were obtained by separately fitting the CTI signal representation to the intracellular and extracellular signals. For cylinders, the intracellular signals give a large negative microscopic kurtosis, while the extracellular space has a small positive microscopic kurtosis. For beads, there is a large positive microscopic kurtosis from the intracellular signals and a very small positive microscopic kurtosis from the extracellular signals.

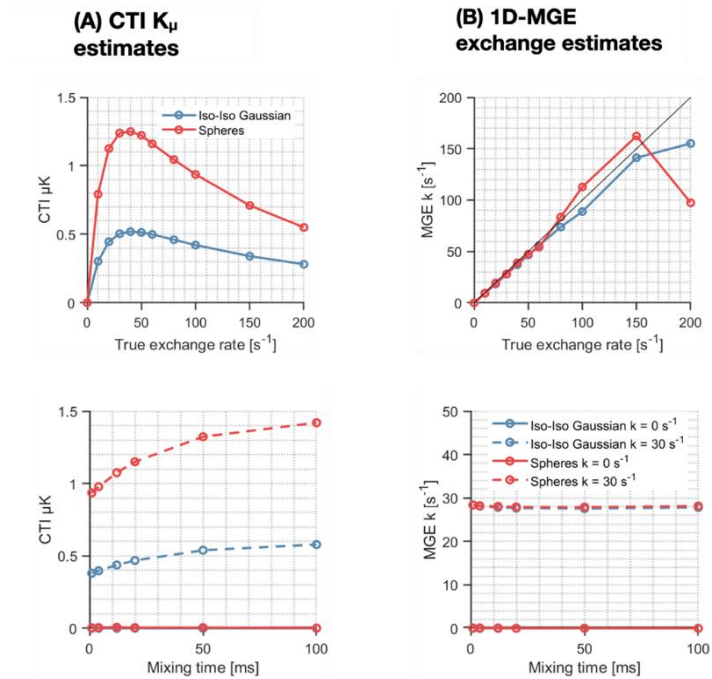

Figure A2: Figure 3 made with the clinical protocol.

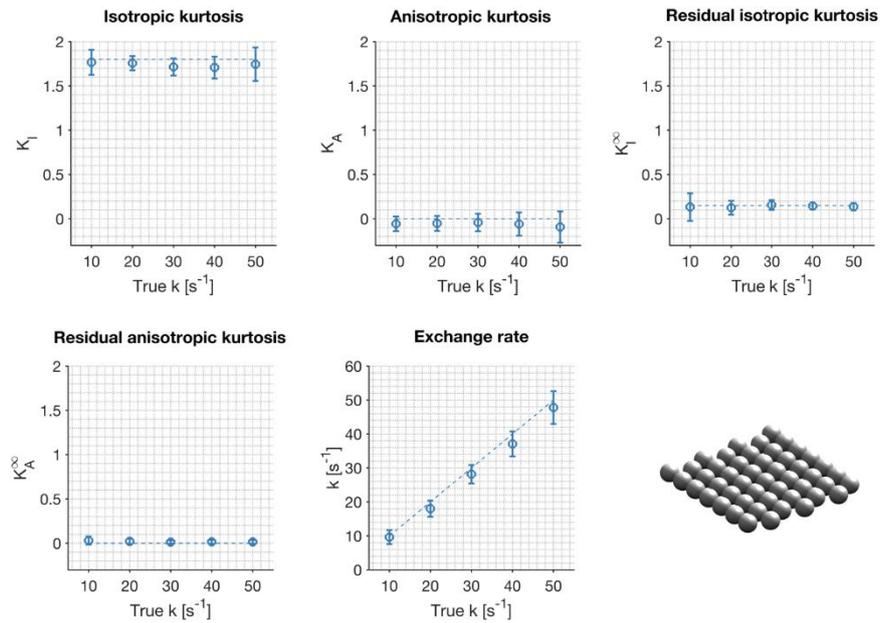

Figure A3: MGE parameter estimates in spheres at SNR = 200. The dashed lines indicate ground truth. The kurtosis is dominated by the isotropic components as expected. All estimates show good agreement with ground truth, with a small bias that can be explained by higher-order effects associated with the cumulant expansion. The error bias represent 1 standard deviation.

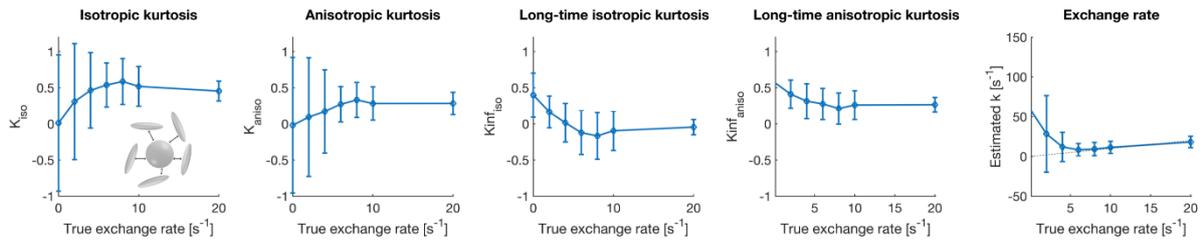

Figure A4: MGE parameter estimates for anisotropic Gaussian components in exchange with an isotropic component. This is the equivalent of Fig. 4 but for lower exchange rates. The dependence on exchange rate of the kurtosis estimates, the bias in estimated exchange rates and the relatively large uncertainties at low exchange rates indicate the difficulty of inverting the MGE signal representation around k = 0.

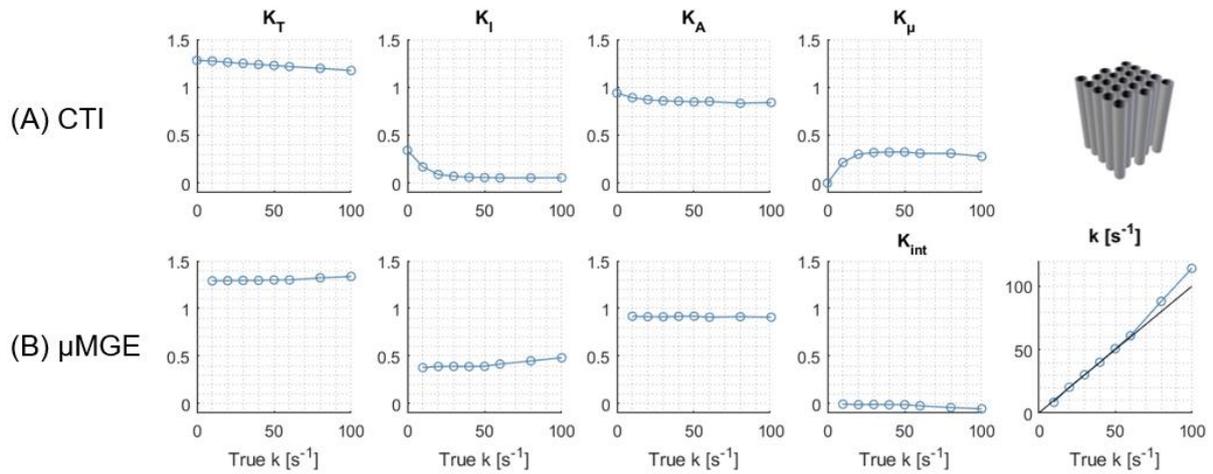

Figure A5: Evaluation of μMGE in a substrate of parallel cylinders. Fitting μMGE allows disentanglement of two sources of microscopic kurtosis: exchange and intrinsic kurtosis.

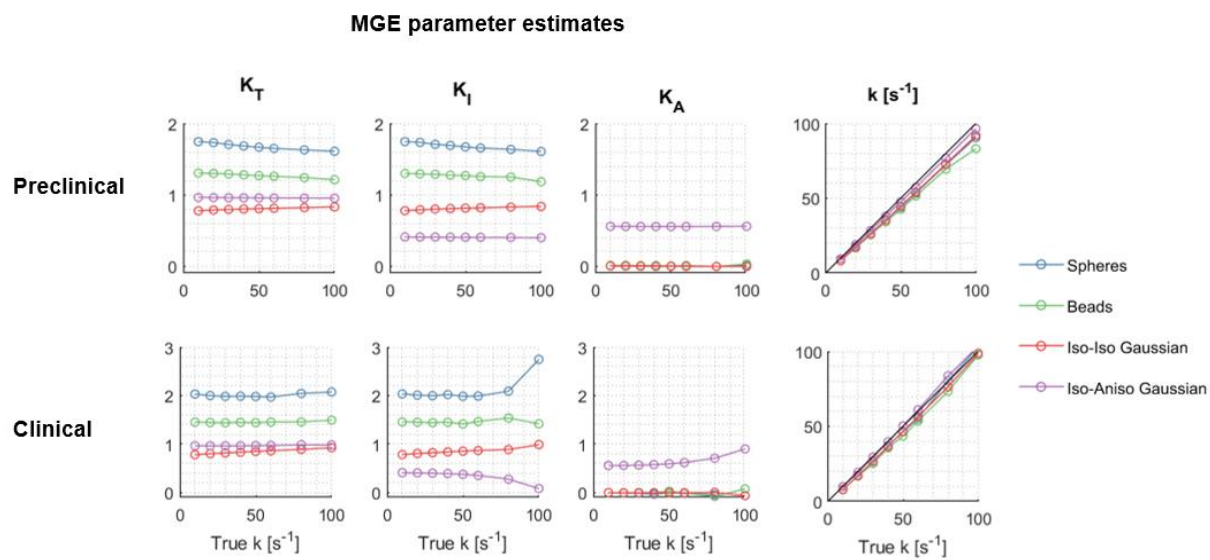

Figure A6: MGE parameter estimates in different substrates with varying rates of exchange. With the preclinical protocol, all kurtosis estimates are largely independent of the underlying exchange rate, and the exchange estimates agree well with the ground truth. Similar trends are observed with the clinical protocol, with the exception of spheres and anisotropic Gaussian where there is a bias in isotropic and anisotropic kurtosis estimates at the fastest exchange rate simulated.